\documentclass[%
 reprint,
 amsmath,amssymb,
aps,
prx,
]{revtex4-1}

\usepackage{graphicx}
\usepackage{dcolumn}
\usepackage{bm}
\usepackage{color}

\begin{document}

\title{Braiding Majorana Zero Mode in An Electrically Controllable Way}
\author{Hai-Yang Ma$^{1}$, Dandan Guan$^{1}$, Shiyong Wang$^{1}$, Yaoyi Li$^{1}$, Canhua Liu$^{1}$, Hao Zheng$^{1}$}
\author{Jin-Feng Jia$^{1,2,3}$}
\email{jfjia@sjtu.edu.cn}
\affiliation{$^{1}$Key Laboratory of Artificial Structures and Quantum Control (Ministry of Education), Shenyang National Laboratory for Materials Science, School of Physics and Astronomy, Shanghai Jiao Tong University, 800 Dongchuan Road, Shanghai 200240, China}
\affiliation{$^{2}$Tsung-Dao Lee Institute, Shanghai 200240, China}
\affiliation{$^3$CAS Center for Excellence in Topological Quantum Computation, University of Chinese Academy of Sciences, Beijing 100190, China}


\begin{abstract}
To realize the braiding operations of Majorana zero mode in the vortex cores of a topological superconductor (TSC), a novel approach is proposed in this letter to replace the common tip (or tip-like) method. Instead of on top of the TSC thin film, arrays of electrically controllable pining centers are built beneath the film, hence detecting can proceed along with braiding. It does not only increase the braiding rate, but also enables the braiding to be performed in an electrically controllable way and to be integrated into large scale. Our work paves the way towards large-scale topological quantum computation.
\end{abstract}

\pacs{Valid PACS appear here}
\maketitle


\section{Introduction}
Quantum computation is believed to be much more powerful than classical one, especially when it comes to  problems which are incapable for a conventional computer, such as the prime factorization \cite{shor1994algorithms,nielsen2002quantum}. However, full-scale quantum computing \cite{preskill2018quantum} is still in its infancy. The challenge lies in its building block, quantum bits (Qubits). Qubits are error-prone, difficult to control during any computation or interaction with the environment, which limits further increasing the number of qubits in a single processor. Then the question is how to build fault-tolerant qubit at the physical level as classic bit in magnetic memory. Topological quantum computation (TQC) is expected to address the issue \cite{kitaev2003fault,freedman2003topological,nayak2008non}. TQC utilizes anyons which has non-Abelian braiding statistics to perform quantum computation \cite{kitaev2003fault,freedman2003topological,nayak2008non}. Qubits in TQC are built non-locally into the quasiparticle states, hence naturally immune to errors caused by local perturbations \cite{kitaev2003fault}. The novel particle in high-energy is called Majorana fermion (MF). In condensed-matter physics, one can alternatively use Majorana zero mode (MZM) sharing similar statistical properties as MF, to build TQC \cite{kitaev2003fault,read2000paired,fu2008superconducting,sau2010generic}. Up to now, MZM has been experimentally observed in the vortex core of artificial topological superconductors (TSC) \cite{xu2014artificial,xu2015experimental,sun2016majorana}, intrinsic TSC \cite{liu2018robust,wang2018evidence,zhang2018observation} and other systems \cite{nadj2014observation}, which accomplished the first step towards TQC, i.e. initialization of MZM. The next step is to perform unitary gate operations for quantum computation, which can be realized with the braiding operations of MZM \cite{nayak2008non} (i.e. move one vortex around another one or more). 

MZM exists in the Abrikosov vortex core of TSC, so braiding MZM is equivalent to braiding the vortices. Abrikosov vortex in a type-II superconductor has two critical length scale: the coherence length of the cooper pairs $\xi$, which determines its size, and the magnetic penetration depth $\lambda$, which is the decay length for the circling supercurrent. Each vortex has a quantized magnetic flux. After a magnetic field is applied, vortices are generated in the superconductor and pinned at the positions where the local potential suppresses the superconductivity (usually defects). Then moving (pushing or dragging) the vortices involves only either depin or repin manipulations \cite{gardner2002manipulation,villegas2003superconducting,de2006controlled,gutierrez2007strong,straver2008controlled,auslaender2009mechanics,kalisky2011behavior,veshchunov2016optical,ge2016nanoscale,kremen2016mechanical,ge2017controlled}. At present, the general idea in experiments to realize braiding is using a tip (or tip-like apparatus) to move the vortices, such as scanning tunneling microscopy (STM) \cite{ge2016nanoscale}, magnetic force microscopy (MFM) \cite{straver2008controlled,auslaender2009mechanics} or scanning superconducting quantum interference device (SQUID) \cite{gardner2002manipulation,kalisky2011behavior,kremen2016mechanical}. For STM, applying tunneling pulse through the tip to the sample would locally heat the sample and suppress the superconductivity at the hot spot \cite{ge2016nanoscale}, then the hot spot serves as a strong pining center to attract the vortices nearby. After re-cooling, the pinned vortex (or vortices) may still be there. Moving the vortex can be accomplished by repeating the heating procedures at the neighboring position. Note that one can also use laser beam to create a local hot spot in the sample \cite{veshchunov2016optical}. For MFM, the tip is magnetic, therefore, the tip can directly interact with the vortex. The tip-vortex interactions can be either repulsive \cite{straver2008controlled} or attractive \cite{auslaender2009mechanics}. By moving the tip, one can either push or drag the vortex to move along with the tip. For scanning SQUID, the tip is pressed into the sample where a stress is locally applied. The superconductivity is then suppressed at the strained position hence attracts the vortex nearby \cite{kremen2016mechanical}. By sweeping the tip (which keeps in touch with the sample) among the sample, the vortex is dragged to move. If the tip of the scanning SQUID is replaced with a magnetic field coil, the scanning SQUID behaves in the same way MFM does \cite{gardner2002manipulation,kalisky2011behavior}. 

However, the above methods have following disadvantages:
\begin{enumerate}
\item[1)] It is very slow to drag or push the vortex with a mechanical-driving tip. And it cannot be expected to speed up since the bindings between a tip and vortex are fragile.
\item[2)] In the tip (or tip-like) way, one cannot directly determine the positions of the vortices since their generations are uncontrollable (note a single vortex-antivortex pair can be generated controllably \cite{ge2017controlled}). Scanning over a large area must be done at first to find the suitable vortices, then can a braiding operation be performed. This uncontrollability is one of the main shortcomings of this method. Moreover, this kind of setup also puts up a difficulty in detection since the tip is occupied by braiding.
\item[3)] The tip (or tip-like) way cannot be integrated into large scale since one tip can only perform one braiding operation at the same time, hence limiting its further developments to full-scale quantum computer.
\end{enumerate}
To overcome these shortcomings, in this letter, we propose an alternative way to perform the braiding. The main idea is to build arrays of electrically controllable pining centers beneath, but not on top of the TSC thin film [Fig. 1]. These pining centers can be built by either local strain, magnetic field or hot spot induced by the tunneling current as mentioned above. After locally suppressing the superconductivity, vortex hosts MZM can be generated and/or pined on top of TSC. By electrically controlling the “moving” of the position of the pining center, a MZM can be moved along a way circling another one or more, which is just the needed braiding operation and this operation should have much faster rate since it is based on controllable electrical circuits. Besides the fast rate, the biggest advantage of our braiding operation method is that it enables the detection while braiding and is easy to integrate into large scale towards industrial devices, which cannot be realized in the common tip way. Furthermore, our proposal is experimentally feasible, since all the needed techniques and requirements are available with nowadays technologies. 

\begin{figure}[tbp]
\includegraphics[width=1.0\linewidth]{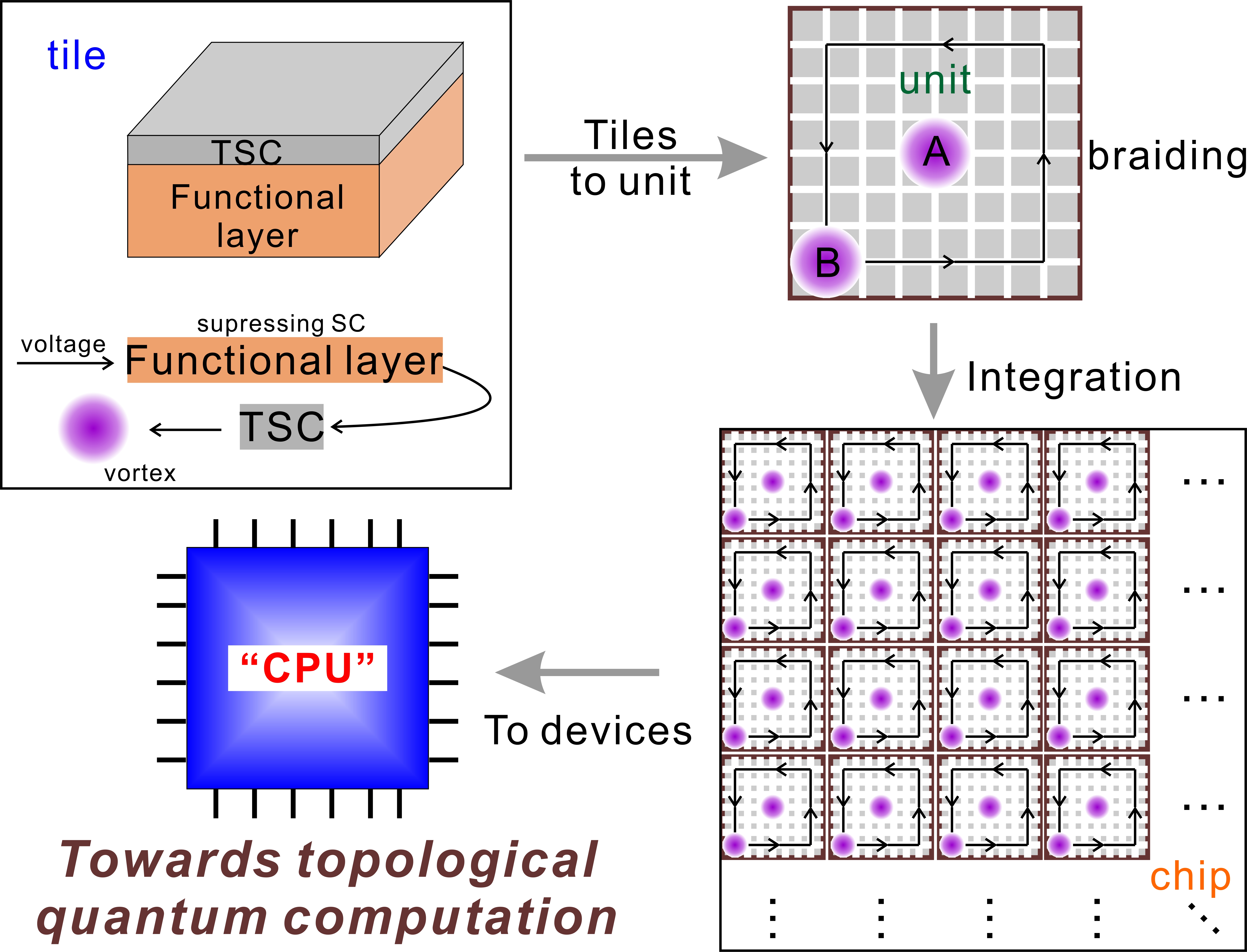}
\caption{\label{fig:1}Schematics of the electronically controllable way towards topological quantum computations. The first step is to construct a tile, which is the basic building block. It is a heterostructure of top topological superconductor (TSC) thin film and bottom functional layer used to suppresses superconductivity. Generations and/or pinning of vortex can be realized by applying a voltage to the functional layer to trigger the pining interactions (usually one need assemble several tiles to build a single pinning center). The second step is to build a braiding unit (Note the TSC layer has not been cut, please see the main text). Braiding of MZM goes as: first, “turn on” the local pining centers of the tiles lying at the center and corner of the unit respectively to generate and/or pin two vortexes A and B. Keep vortex A intact, gradually remove the voltage of the tiles right down the vortex B, at the same time, gradually turn on the voltage of the nearby tiles, then we succeed in moving MZM by one tile distance. Repeat the procedures and moving vortex B along a way circling vortex A, then we finish one desired braiding operation of the MZM. Finally, integrate the braiding units into large-scale “chip” and seal them into “CPU” devices towards full-scale topological quantum computation.}
\end{figure}

\section{Proposals and discussions}
Here take building pinning centers with local strain as an example, we give a full description of this method in the following text while the other two methods, namely building pinning centers with local magnetic field or local hot spot induced by a tunneling current, will be discussed later. 

Our proposal towards braiding MZM in terms of local strain proceeds in four steps. As shown in Fig. 1, the first step is to construct a single tile. The tiles are heterostructures of the top TSC thin film and bottom piezoelectric material (PZT) layer [Fig. 2(a)]. PZT materials such as piezoceramics have the property that they can transform electrical voltage into stress, which is essential for the electrical controllability of the local strain. The pinning of a vortex on the TSC then goes as follows [Fig. 2(a)]: first apply an electrical voltage on PZT layer; the PZT layer then transform the electrical voltage into stress which will generate finite strain on the TSC above, this strain may locally suppress the superconductivity and finally contribute to pinning vortex27. The first step is highly feasible for experimental setup and may be achieved in the near future.

With tiles at hand, the second step is to assembly some of them into braiding unit, as shown in the top-right panel of Fig. 1. The size of the braiding unit needs to be large enough to accommodate two vortices. Then the braiding of MZM can be accomplished in a series of operations: First, turn on the local strain of the tiles lie at the center and corner of the unit respectively to pin two vortices A and B (top-right panel of Fig. 1). Keeps vortex A at the center intact, gradually remove the local strain of the tiles right down the vortex B at the corner. At the same time, gradually turn on the local strain of the nearby tiles of vortex B, then we succeed in moving the pinning center hence the MZM by one tile distance. Repeat the above procedures and move vortex B along a way circling vortex A, then one desired braiding operation of the MZM is completed. More complicated braiding operations can start with this and includes more units and operations. These operations require properly programmed electric circuits, which are accessible with current technologies.

Step 2 is a benchmark of our proposal towards TQC. Once succeeds, step 3 and step 4 then are straightforward. Just by integrating more and more braiding units, we obtain an electrically based “chip” and finally industrial “CPU” which may be able to perform full-scale TQC, see the bottom panels of Fig. 1. However, in real fabrication of the “chip”, the procedures go in the opposite direction as from step 3 to step 1. That is why we prepare a large plate of PZT layers first, and then “cut” them into meshes of tiles using lithography; after setting up all the electric circuits, deposit a thin film of TSC on top of the PZT layer (whether the TSC layer need cutting or not depends on practical requisite), then we get a “chip” of tiles. As we can see, our proposal is originally feasible towards large-scale TQC.

Note for the above toy buildings, we have ignored many details. Here we comment on two of the most important requirements that must be fulfilled in the experimental implementation:
\begin{enumerate}
\item[i)] The size of the tiles should be smaller than the size of the vortex (depends on the coherence length $\xi$), otherwise we cannot finely generate and control the pinning and moving of the vortices, which will lead to inevitable failure of the braiding operations described above. Typically, the size of the tiles can be as large as $\sim$40 nm \cite{xu2014artificial,xu2015experimental}, which is an easy task for current lithography technique.
\item[ii)] The local strain may not be able to fully suppress the local superconductivity of the TSC, therefore, it can only serve as a pining center. Then a question is: how can we initialize a vortex at the pinning center? This difficulty can be solved by first triggering local strain at specific positions as artificial pining centers, then applying suitable magnetic field to generate vortices which may be pinned by the established pining centers. This idea can be applied to the other two methods as well. But anyway, such a requirement put a limitation on real implementations of this method.
\end{enumerate}

\begin{figure}[tbp]
\includegraphics[width=1.0\linewidth]{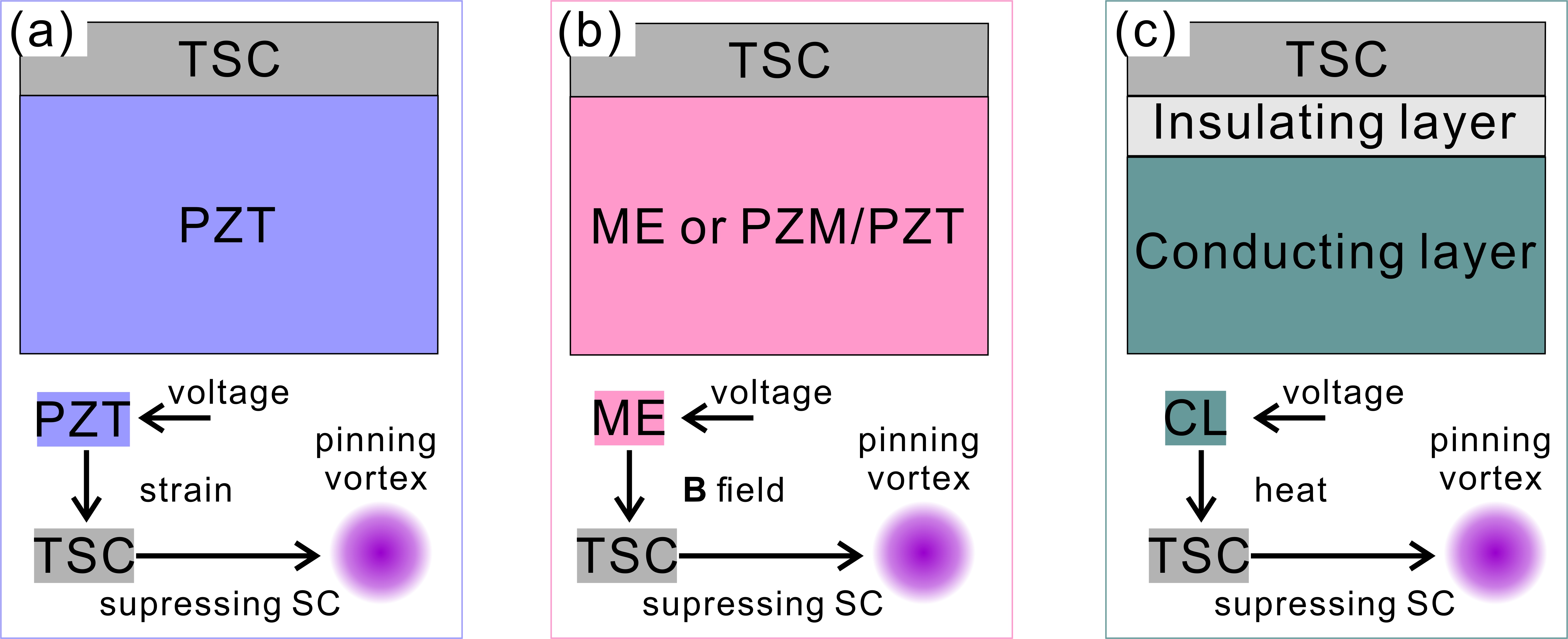}
\caption{\label{fig:2}Schematics of different implementations of the local pinning centers. (a) Local pinning center built with piezomagnetic materials (PZT). The pinning of a vortex on the TSC then goes as follows: first apply an electrical voltage on PZT layer; the PZT layer then transforms the electrical voltage into stress which will generate a strain on the TSC above, this strain may locally suppress the superconductivity (SC) and contribute to pinning vortex. (b) Local pinning center built with magnetoelectrical (ME) materials. The ME layer may be intrinsic ME material or equivalent ME materials such PZM/PZT heterostructures. The generation and/or pin a vortex goes as follows: first apply electrical field on ME layer; the ME layer then transforms the electrical field into magnetic field and effects on the TSC above, which contributes to generating and/or pinning vortex. (c) Local pinning center built with thin insulating layer and conducting layer. The pinning of a vortex on the TSC goes as follows: first apply an electrical voltage on conducting layer (CL); then a locally tunneling current is generated between the TSC and the conducing layer, this tunneling current will locally heat the TSC which serves as a hot spot and locally suppress the superconductivity and contribute to pinning vortex.}
\end{figure}

If alternatively using electrically controllable local magnetic field to build the pinning centers, the limitation can be removed since the generation of vortex is just the generation of magnetic flux. The tiles now are heterostructures of the top TSC thin film and bottom magnetoelectric (ME), or equivalent MZ layer [Fig. 2(b)]. ME material has the property that it can transform applied electrical filed into magnetic field, which is essential for the electrical controllability of the local magnetic fields. The generation and/or pin a vortex goes as follows [Fig. 2(b)]: first apply electrical field on ME layer; the ME layer then transforms the electrical field into magnetic field which effects on the TSC above, contributes to generating and/or pinning vortex. However, the induced local magnetic fields should be strong enough to generate and/or pin a vortex in the TSC. In fact, on the one hand, the required strength of the magnetic field is quite small for some artificially topological superconductors \cite{sun2016majorana}. On the other hand, this requirement can be relieved by applying an external magnetic field on the sample to help generate the vortex. The needed strength of the superposed local filed to generated and/or pin and move the vortex therefore can be quite small and should be approachable for ME (or equivalent ME) materials. The main goal then is to search for materials with giant ME effect. Usually the ME effect is weak for intrinsic ME materials (eg. Cr$_{2}$O$_{3}$ \cite{astrov1961magnetoelectric}) and one may refer to multiferroic material such as Y-type hexaferrite Ba$_{2-x}$Sr$_{x}$Mg$_{2}$Fe$_{12}$O$_{22}$ family \cite{chun2010realization,zhai2017giant}, $R$MnO$_{3}$ ($R$ = Dy, Tb, and Gd) \cite{aoyama2015multiferroicity}, GdMn$_{2}$O$_{5}$ \cite{lee2013giant} etc., to reach large ME effects \cite{parkes2012non,eerenstein2006multiferroic,binek2004magnetoelectronics}. Alternatively, we can also build equivalent ME effect with heterostructures, for example by putting a piezomagnetic material \cite{jaime2017piezomagnetism,boldrin2018giant,mahaiyang2020svl} (PZM) and a piezoelectric ceramic together \cite{fiebig2005revival}. For the PZM/PZT heterostructure, the equivalent ME effect is generated from electric filed to strain field in the PZT layer due to piezoelectric effect, and then strain field to net magnetizations in the PZM layer through piezomagnetic effect. In this way, we can build electrically controllable local magnetic field. With tiles being prepared, steps 2 to 4 are just the same as before.

Other than the two methods, one can also use electrically controllable hot spot to build the pining centers \cite{veshchunov2016optical,ge2016nanoscale}. The bottom layer of the heterostructures here is a little bit more complicated, see Fig. 2(c). Between the top TSC thin layer and the beneath conducting layer, there is a thin insulating layer. This insulating layer helps to generate a tunneling current and should be thin enough so that a finite tunneling current from the TSC layer to the conducting layer can be generated upon and apply a suitable voltage on the conducting layer. The pinning of a vortex on the TSC then goes as follows [see Fig. 2(c)]: first apply an electrical voltage on conducting layer; then a locally tunneling current is generated between the TSC and the conducing layer, this tunneling current will locally heat the TSC which serves as a hot spot that locally suppress the superconductivity and contribute to pinning vortex. If the heating is strong enough, locally the superconductivity may be fully suppressed, then vortex or vortices can be generated and keep there even after removing the voltage \cite{ge2016nanoscale}. Therefore, one must carefully tune the voltage to make sure that only one vortex is pinned. But as pointed out above, we may need not finely tune the voltage to initialize a vortex. Initializing of the vortices can be accomplished by simply applying a small voltage to building pinning centers first and then apply magnetic field to generate vortices which may be pinned the established pinning centers. Moving the vortex also has no need for high voltages, making it more achievable for experimental implementation. With tiles being prepared, steps 2 to 4 are just the same as before.

\section{Conclusions}
To conclude, we proposed an alternative way other than the common tip method to realize the braiding operations of MZM. There are some special advantages for our proposals. First, the vortices on top of TSC are electrically controllable, giving rise to much faster braiding rate and easy manipulation. Second, the arrays of electrically controllable pining centers are built beneath the film, which enables the detection while braiding. Third, the braiding units of our proposal can be easily integrated into large scale. This property is of great importance in developing full-scale TQC and one sees no such possibility with the common tip way before. However, we are aware that the proposals in this letter are difficult for real implementations. But all the requirements and needed techniques are accessible with nowadays technologies, in principle, the proposed methods for the braiding operation of MZM are expected to be feasible. Our proposals may open a new era for the experimental implementation of TQC.

\begin{acknowledgments}
We thank the Ministry of Science and Technology of China (Grants No. 2019YFA0308600, 2020YFA0309000, 2016YFA0301003, No. 2016YFA0300403), NSFC(Grants No. 11521404, No. 11634009, No. 11874256, No. 11874258, No. 12074247, No. 11790313, and No. 11861161003), the Strategic Priority Research Program of Chinese Academy of Sciences (Grant No. XDB28000000) and the Science and Technology Commission of Shanghai Municipality (Grants No. 2019SHZDZX01, No. 19JC1412701, No. 20QA1405100) for partial support.
\end{acknowledgments}

\nocite{*}

\bibliography{BraidingMZM}

\end{document}